\documentclass[12pt,preprint]{aastex}
\def\({\left(}
\def\){\right)}
\def\[{\left[}
\def\]{\right]}

\usepackage{graphicx}
\usepackage{natbib}
\usepackage[usenames,dvips]{color}

\begin{document}
\title{Application of the Disk Evaporation Model to AGNs}

\author{B. F. Liu\altaffilmark{1}
\email{bfliu@ynao.ac.cn}
\and
Ronald E. Taam\altaffilmark{2,3,4}
\email{r-taam@northwestern.edu}}
\altaffiltext{1}{National Astronomical Observatories/Yunnan Observatory,
Chinese Academy of Sciences, P.O. Box 110, Kunming 650011, China}
\altaffiltext{2}{Academia Sinica Institute of Astrophysics and Astronomy-TIARA,
P.O. Box 23-141, Taipei, 10617 Taiwan}
\altaffiltext{3}{Academia Sinica Institute of Astrophysics and Astronomy/National
Tsing Hua University-TIARA, Hsinchu Taiwan}
\altaffiltext{4}{Northwestern University, Department of Physics and Astronomy,
2131 Tech Drive, Evanston, IL 60208}

\begin{abstract}
The disk corona evaporation model extensively developed for the interpretation of
observational features of black hole X-ray binaries (BHXRBs) is applied to AGNs.
Since the evaporation of gas in the disk can lead to its truncation for accretion
rates less than a maximal evaporation rate, the model can naturally account for the
soft spectrum in high luminosity AGNs and the hard spectrum in low luminosity AGNs.
The existence of two different luminosity levels describing transitions
from the soft to hard state and from the hard to soft state in BHXRBs, when applied
to AGNs, suggests that AGNs can be in either spectral state within a range of
luminosities. For example, at a viscosity parameter, $\alpha$, equal to 0.3, the
Eddington ratio from the hard to soft transition and from the soft to hard transition
occurs at 0.027 and 0.005 respectively. The differing Eddington ratios result from
the importance of Compton cooling in the latter transition, in which the cooling
associated with soft photons emitted by the optically thick inner disk in the soft
spectral state inhibits evaporation.  When the Eddington ratio of the AGN lies below
the critical value corresponding to its evolutionary state, the disk is truncated. With
decreasing Eddington ratios, the inner edge of the disk increases to greater distances
from the black hole with a concomitant increase in the inner radius of the broad line region,
$R_{BLR}$. The absence of an optically thick inner disk at low luminosities ($L$) 
gives rise to region in the
$R_{BLR}-L$ plane for which the relation $R_{BLR}
\propto L^{1/2}$ inferred at high luminosities is excluded. As a result, a lower limit
to the accretion rate is predicted for the observability of broad emission lines, if the
broad line region is associated with an optically thick accretion disk.  Thus, true
Seyfert 2 galaxies may exist at very low accretion rates/luminosities. The differences
between BHXRBs and AGNs in the framework of the disk corona model are
discussed and possible modifications to the model are briefly suggested.
\end{abstract}

\section{Introduction}
The structure of accretion disks surrounding compact objects is central to our understanding
of both stellar mass black holes in X-ray binary systems and supermassive black holes in the
nuclei of galaxies.  Among the issues requiring elucidation is the physical description of
the accretion flow in the sources' spectral states. Toward this end, Shakura \& Sunyeav (1973)
pioneered the development of optically thick and geometrically thin accretion disks within the
framework of a turbulent $\alpha$ disk model, which provided an explanation of the thermal state
of these sources.  Recognizing the possibility of physically distinct states in the disk to explain
the hard X-ray state of Cyg X-1 (e.g., Agrawal et al. 1972; Tananbaum et al. 1972), Shapiro,
Lightman, \& Eardley (1976) and Ichimaru (1977) proposed a geometrically thick and optically thin
accretion flow model. More recently, Narayan \& Yi (1994, 1995a, 1995b) and Abramowicz et al.
(1995) have stressed the importance of the advection of accretion energy stored in the accreting
gas, leading to radiative inefficient accretion in the so-called advection-dominated accretion
flow (ADAF).  The investigation of the features of these ADAF models and their application to
the spectral states of black hole X-ray binary systems and AGNs has been discussed in
the reviews of Narayan, Mahadevan, \& Quataert (1998), Narayan (2005), and Narayan \&
McClintock (2008).

The high/soft state spectrum of BHXRBs is generally believed to originate from an accretion disk
extending to the innermost stable circular orbit (ISCO) as developed by Shakura \& Sunyaev (1973),
whereas the low/hard state spectrum is thought to originate in a geometrically thick, optically
thin, hot corona or ADAF in the immediate vicinity of the black hole (see Narayan \& Yi 1994;
1995a, b). We note that the observational data of accreting black holes commonly point to the
coexistence of hot and cool accretion flows, which is likely to be either in the form of an inner
ADAF connecting to an outer geometrically thin disk or an ADAF-like corona lying above a standard
thin disk extending to the ISCO. The former configuration has been studied by Esin et al. (1997)
and applied to both quiescent BHXRBs and low-luminosity AGNs.  In contrast, the latter configuration was
proposed to explain the power-law X-ray emission and lower-frequency blackbody component observed
in both BHXRBs and AGNs (e.g., see Liang \& Price 1977; Haardt \& Maraschi 1991; Nakamura
\& Osaki 1993). In the context of these studies, only the disk evaporation model (e.g. Meyer et
al. 2000a) has been investigated in detail to elucidate the nature of the interaction between
the disk and the corona.

Such an evaporation model has been reasonably successful in providing an interpretive
framework of the many features observed in BHXRBs. For example, the spectral transition
between the low/hard and high/soft state (Meyer et al. 2000b), the disk truncation and
formation of an ADAF in the low/hard state (Liu et al. 1999;
Meyer-Hofmeister \& Meyer 2003), the luminosity hysteresis between the hard-to-soft and
soft-to-hard transitions (Meyer-Hofmeister et al.  2005; Liu et al. 2005), the intermediate
state (Liu et al. 2006; Meyer et al. 2007), and the inner cool disk component in the low/hard
state (Liu et al. 2007; Taam et al.  2008) can all be understood as consequences of the disk
evaporation/condensation process. The model provides a physical basis for understanding the
observational phenomenology, although details remain to be addressed.

As is well known, observations of AGNs exhibit many similar features to BHXRBs. Of importance is
the existence of the fundamental plane (Merloni et al. 2003; Falcke et al. 2004; Wang et al. 2006) relating the radio
luminosity, X-ray luminosity and black hole mass which extends from stellar masses to supermassive
black holes.  In addition to these relations, the differing spectral shapes between high luminosity
AGNs (HLAGNs) and low luminosity AGNs (LLAGNs) (e.g., Elvis et al. 1994; Ho 1999; Ho 2008 and reference
therein) are widely observed.  Specifically, the HLAGNs, such as quasars and some Seyfert galaxies,
display evidence of disk accretion as inferred from the presence of a ``big blue bump'' and hence
are the analogues of high/soft state BHXRBs (Maccarone et al. 2003). On the other hand, LLAGNs
exhibit hard X-ray emission as illustrated, for example, by our Galactic Center (Narayan
\& Yi 1994; Yuan et al.  2003; Narayan 2005) and/or evidence of radio jets and are analogues
of the low/hard state BHXRBs
(for a review see Ho 2008). Spectral features seen in narrow line Seyfert 1 galaxies (NLS1) are
similar to BHXRBs at the very high states (e.g. Pounds et al. 1995; Pounds \& Vaughan
2000). Although there exist differences (e.g., a soft X-ray excess is often seen in
AGNs), a phenomenology based on a unification scheme for the BHXRBs and AGNs has been proposed
(Falcke et al.  2004; Fender et al. 2004; Ho 2009).

In these models, the transition from an outer disk to an inner ADAF is one of the primary issues
remaining to be clarified.  Among the mechanisms facilitating the state transition (see Honma 1996;
Manmoto \& Kato 2000; Meyer et al. 2000b; R\`o\.za\`nska \& Czerny 2000a,b; Spruit \& Deufel 2002;
Lu et al. 2004; Dullemond \& Spruit 2005), the disk corona evaporation model provides a promising
explanation for not only the state transition, but also the observational features of BHXRBs
aforementioned.  It is, thus, natural to confront the disk corona evaporation model to observations
of AGNs to determine its applicability to accreting black holes systems in general.

Investigation of the model reveals that the evaporation feature is independent of
the black hole mass. In particular, the coronal temperatures in the inner region of
disks around supermassive black holes and in BHXRBs are the same.  Furthermore, the radial
distribution of the evaporation rate and the mass flow rate when scaled to the black hole
mass for both AGNs and BHXRBs are also independent of the mass. Thus, the model can, in
principle, be applied to AGNs (Liu \& Meyer-Hofmeister 2001), especially given the
analogues between the observational
features of AGNs and BHXRBs.  In this regard, a spectral state transition for AGNs with a
luminosity of a few percent of the Eddington value, the truncation of a thin disk for
LLAGNs, the presence of iron lines related with both the disk and the corona, and the occurrence of
radio jets associated with the evaporative process in disk coronal flows may be expected.

To investigate the disk evaporation/condensation model for AGNs, we outline the basic physics
and assumptions of the theoretical model in \S 2 and discuss its applicability as an interpretative
framework for several AGN properties in \S 3. The implications of the model and the issues that
remain to be addressed due to differences between AGNs and BHXRBs are presented in \S 4.

\section{Theoretical Model}
The concept of the evaporation of matter via a siphon flow from an accretion disk was
originally proposed in the pioneering work by Meyer \& Meyer-Hofmeister (1994) to explain
the enigmatic phenomenon of the UV lag observed in dwarf novae. Here, the evaporation leads
to the formation of a hole in the inner region of an accretion disk, resulting in the
retardation of the UV radiation during the onset of an outburst. The model was developed
in a more detailed form for application to disks surrounding black holes by Meyer et al.
(2000a), who demonstrated the possible truncation of an outer optically thick disk and the
consequent formation of an inner optically thin ADAF. Liu et al. (2002a) extended the model
to take account of the decoupling of ions and electrons and the effect of Compton scattering
in the inner regions of the disk close to the central black hole. Similar conceptual models,
but with differences in detail were advanced in a semi-analytical model proposed by
Dullemond \& Spruit (2005) and in a vertically stratified model developed by R\`o\.za\`nska
\& Czerny (2000a,b).  The incorporation of inflow and outflow of mass, energy and angular
momentum from and towards neighboring zones was included by Meyer-Hofmeister \& Meyer
(2003), and the effect of magnetic fields with consequentially reduced conduction efficiency
was investigated by Meyer \& Meyer-Hofmeister (2002), Meyer-Hofmeister \& Meyer (2006),
and Qian, Liu, \& Wu (2007).  More recently, the model has been further developed to include
the condensation of coronal gas to an inner disk (Meyer et al. 2007; Liu et al. 2007;
Taam et al. 2008).  The basic physics and assumptions of the model are described in the
following.

\subsection{Model assumptions}

We investigate stationary accretion to a non-rotating black hole. The accretion takes place
via a standard optically thick and geometrically thin disk (Shakura \& Sunyaev 1973)
enclosed by a hot, accreting corona. Such a corona may form by processes similar to those
operating in the surface of the sun, or from a thermal instability in the uppermost layers
of the disk (e.g. Shaviv \& Wehrse 1986). Both the disk and corona are individually powered
by the release of gravitational energy associated with the accretion of matter affected
through viscous stresses. The interaction between the corona and the underlying disk results
from the vertical conduction of heat by electrons, mass evaporation/condensation and hence
enthalpy flow, and inverse Compton scattering of the disk photons by coronal electrons. The
disk evaporation/condensation model takes account of both the mass exchange (mass
evaporation/condensation) and energy exchange (conduction and enthalpy flux) between the
disk and corona.  This is in contrast to classical disk corona models where only
radiative coupling is considered (e.g. Haardt \& Maraschi 1991). Assuming a radial one
zone approximation, we investigate the mass flow and energy balance in the corona and
calculate the detailed vertical structure and radiation emitted from the corona and disk
as functions of the viscosity parameter in the corona, black hole mass, and
mass accretion rate.

\subsection{General description of evaporation}
For the modeling of the accretion process, the interaction between the disk and corona is an
important and distinctive process compared to the case for an ADAF or disk alone.  The corona is
an ADAF-like accretion flow modified by the vertical heat conduction and inverse Compton
scattering of disk photons, which play a key role in cooling the electrons as a consequence
of the soft photon flux from the underlying disk.  The ions in the corona are directly
heated by viscous dissipation, partially transferring their energy to the electrons by
means of Coulomb collisions. The energy gained by electrons cannot be effectively radiated
away and is conducted to the lower, cooler, and denser coronal layers by electron-electron
collisions.  In the transition layer, the conductive heat flux is radiated away through
bremsstrahlung only if the number density in this layer reaches a critical value.  If the
density is too low to efficiently radiate the energy (lower than the critical value), a
certain amount of lower, cooler gas is heated up,  whereby the bremsstrahlung cooling
rate is raised, until an energy equilibrium is established between the conduction,
radiation and heating of the cool gas. The transfer of gas from the disk to the corona,
to establish the equilibrium, is called evaporation. It occurs in the low density
coronae. If the density is so high that bremsstrahlung is more
efficient than the conductive heating, a certain amount of gas is over-cooled and
condenses onto the disk until energy equilibrium is reached.  This condensation process
takes place in the high-density corona. Efficient cooling processes such as strong inverse
Compton scattering of the disk photons can also facilitate such condensation.

The gas evaporating into the corona retains angular momentum and differentially rotates around
the central object. By frictional stresses, the gas loses angular momentum and drifts inward
in such a way that the corona continuously drains gas towards the central object. The coronal
accretion flow is re-supplied by continuous evaporation and, therefore, steady flows are
established in the disk and corona with the mass exchange between the two flows.

In the disk corona model, we assume that the accreting gas is cold and thus enters the disk
rather than a corona as occurs in a binary system where the mass transferred from the donor
star is constrained to the orbital plane. That is, we do not consider the possibility
that the accreting gas enters the disk as a corona (see discussion in sect.4.2) , which may be appropriate for the Galactic
Center or for very low luminosity AGNs. Hence, the model is restricted to systems in the
low/hard state and not necessarily to the very low quiescent state.  Within our approximation, the
coronal accretion is supplied by evaporating disk matter at distances remote from the
accreting black hole.  The mass flow rate through the corona at a given distance is the sum
of the mass evaporation from the outer edge of the disk inward to a given distance.
Hence the mass flow rate in the corona
increases toward the black hole as a result of the radial contribution to the mass
evaporation, notwithstanding the possible loss of gas due to disk outflows.  Note that it is
possible that the mass flow rate contributed by evaporation in the outer region is sufficiently
high that evaporation ceases at some radius and condensation takes place in the inner region.

In the following, the equations which include the detailed physical processes described above
are given.  The model used here is based on Liu et al. (2002a) and includes modifications
(Meyer-Hofmeister \& Meyer 2003) and updates incorporated in recent years (e.g. Qian, Liu, \& Wu
2007; Qiao \& Liu 2009).

Equation of state
\begin{equation}\label{e:EOS}
P={\Re \rho \over 2\mu} (T_{i}+T_{e}).
\end{equation}

Equation of continuity
\begin{equation}\label{e:continuity}
\centering
 {d\over dz}(\rho v_z)=\eta_M{2\over R}\rho v_R -{2z\over
R^2+z^2}\rho v_z.
\end{equation}

Equation of the $z$-component of momentum
\begin{equation}\label{e:mdot}
\rho v_z {dv_z\over dz}=-{dP\over dz}-\rho {GMz\over
(R^2+z^2)^{3/2}}.
\end{equation}

The energy equation of the ions
\begin{equation}\label{e:energy-i}
\begin{array}{l}
{d\over dz}\left\{\rho_i v_z \left[{v^2\over 2}+{\gamma\over
\gamma-1}{P_i\over \rho_i}-{GM\over (R^2+z^2)^{1\over
2}}\right]\right\}\\
={3\over 2}\alpha P\Omega-q_{ie}\\
+{\eta_E}{2\over R}\rho_i v_R
\left[{v^2\over 2}+{\gamma\over \gamma-1}{P_i\over \rho_i}-{GM\over (R^2+z^2)^{1\over 2}}\right]\\
-{2z\over {R^2+z^2}}\left\{\rho_i v_z \left[{v^2\over 2}+{\gamma\over
\gamma-1}{P_i\over \rho_i}-{GM\over (R^2+z^2)^{1\over 2}}\right]\right\},
\end{array}
\end{equation}
where  $q_{ie}$ is the energy exchange rate between the electrons and the ions,
\begin{equation}
{q_{ie}}={\bigg({2\over \pi}\bigg)}^{1\over 2}{3\over 2}{m_e\over
m_p}{\ln\Lambda}{\sigma_T c n_e n_i}(\kappa T_i-\kappa T_e)
{{1+{T_*}^{1\over 2}}\over {{T_*}^{3\over 2}}}
\end{equation}
with
\begin{equation}
T_*={{\kappa T_e}\over{m_e c^2}}\bigg(1+{m_e\over m_p}{T_i\over
T_e}\bigg).
\end{equation}

The energy equation for both the ions and the electrons is
\begin{equation}\label{e:energy-t}
\begin{array}{l}
{\frac{d}{dz}\left\{\rho {v}_z\left[{v^2\over
2}+{\gamma\over\gamma-1}{P\over\rho}
-{GM\over\left(R^2+z^2\right)^{1/2}}\right]
 + F_c \right\}}\\
=\frac{3}{2}\alpha P{\mit\Omega}-n_{e}n_{i}L(T_e)-q_{\rm Comp}\\
+\eta_E{2\over R}\rho v_R \left[{v^2\over
2}+{\gamma\over\gamma-1}{P\over\rho}
-{GM\over\left(R^2+z^2\right]^{1/2}}\right]\\
-{2z\over R^2+z^2}\left\{\rho v_z\left[{v^2\over
2}+{\gamma\over\gamma-1}{P\over\rho}-
{GM\over\left(R^2+z^2\right)^{1/2}}\right]
+F_c\right\},
\end{array}
\end{equation}
where $n_{e}n_{i}L(T_e)$ is the bremsstrahlung cooling rate, of which the radiative cooling function
$L(T_e)$  is taken  from Raymond et al. (1976). Here, ${q_{\rm {Comp}}}$  is the Compton cooling rate,
\begin{equation}\label{e:comp}
 {q_{\rm Comp}}={4\kappa T_e\over m_e
c^2}n_e\sigma_T c u,
\end{equation}
with $u$ the energy density of the soft photon field.  The thermal conduction flux,
$F_c$, is given by (Spitzer 1962)
\begin{equation}\label{e:fc}
F_c=-\kappa_0T_e^{5/2}{dT_e\over dz}
\end{equation}
with $\kappa_0 = 10^{-6}{\rm erg\,s^{-1}cm^{-1}K^{-7/2}}$ for a fully ionized plasma.

All parameters in the above equations are in cgs units and are defined as follows.
Specifically, $P,\rho,T_i$ and $T_e$ are the pressure, density, ion temperature and
electron temperature respectively.  The vertical and radial speed of the mass flow are
denoted by $v_z$ and $v_R$.  In Eq.(\ref{e:EOS}), $\mu=0.62$ is the mean molecular
weight assuming a standard chemical composition ($X=0.75, Y=0.25$) for the corona. The
number density of the ions, $n_i$, is assumed equal to that of the electrons, $n_e$, for
convenience, which is strictly true only for a pure hydrogen plasma.  In the continuity
and energy equations, the terms starting with $2z/(R^2+z^2)$ derive from
the gradual expansion of the vertical flow channel with height as the geometry changes
from cylindrical to spherical, and $\eta_M$ is the mass advection modification term and
$\eta_E$ is the energy modification term.  These terms parameterize the difference between
lateral inflows and outflows. We take $\eta_M=1$ for the case without consideration of
the effect of mass inflow from the outer neighboring zone of the corona, and
$\eta_E=\eta_M+0.5$ is a modification to the previous energy equations (for details see
Meyer-Hofmeister \& Meyer 2003).  The other quantities are as follows: $G$ is the
gravitational constant, $M$ the mass of the accreting black hole, $m_p$ and $m_e$ are respectively
the mass of the proton and the electron, $\kappa$ is the Boltzmann constant, $c$ the speed of light,
$a$ the radiation constant, $\sigma$ Stefan-Boltzmann constant, $\sigma_T$ the Thomson scattering
cross section, $\gamma=5/3$ is the ratio of specific heats, and $\ln\Lambda=20$ is the Coulomb
logarithm.

The five differential equations, Eqs.(\ref{e:continuity}), (\ref{e:mdot}), (\ref{e:energy-i}),
(\ref{e:energy-t}), and (\ref{e:fc}), which contain five variables $P(z)$, $T_i(z)$, $T_e(z)$ , $F_c(z)$, and
$\dot m (z)(\equiv \rho (z) v_z)$, can be solved with five boundary conditions.  In particular, at
the lower boundary of the interface between the disk and corona, the conductive flux
is exactly radiated away and there is no downward heat flux. The temperature of the gas should
be the effective temperature of the accretion disk.  Previous investigations (Liu, Meyer, \&
Meyer-Hofmeister 1995) show that the coronal temperature increases from the effective temperature to $10^{6.5}$K in a very thin layer and that the conductive flux can be expressed as a function
of pressure at this temperature. Thus, the lower boundary conditions can be reasonably
approximated (Meyer et al.  2000a) as
\begin{equation}
T_i=T_e=10^{6.5}K,\ {\rm and} \  F_c=-2.73\times 10^6 P\ {\rm at}\
z=z_0.
\end{equation}
At infinity, there is no artificial confinement and hence no pressure. This requires
a sonic transition at some height $z=z_1$. As there is no heat flux from/to infinity
we constrain the upper boundary as,
\begin{equation}
F_c=0\  {\rm and}
 \  v_z^2=V_s^2\equiv P/\rho={\Re\over 2\mu}(T_i+T_e)\  {\rm at}\  z=z_1.
\end{equation}
With the above boundary conditions, we assume a set of trial lower boundary values for $P$ and $\dot m$ to
start the integration along $z$. Only when the trial values for $P$ and $\dot m$ fulfill the upper
boundary conditions can the presumed $P$ and $\dot m$ be taken as true solutions of the differential
equations.

\subsection{Caveats}
In order to reveal the basic properties of disk corona in AGNs, we make some simplifications and
approximations. Specifically, we assume an $\alpha$ prescription for the viscosity. In reality,
the disk corona structure depends on the magnitude and functional form of the viscosity. This
is likely due to magnetic effects associated with the magnetorotational instability (see Balbus
\& Hawley 1991), though its effectiveness in providing a sufficiently large $\alpha$ remains to
be understood (see King, Pringle \& Livio 2007). The magnetic field also influences the rate
of heating (via non-ideal effects associated with magnetic reconnection) and cooling of the
system since non-thermal radiation processes associated with
synchrotron radiation and synchrotron self-Compton emission could be operating. In addition,
the field can reduce the effectiveness of the vertical heat conduction. Thus, the field likely
plays an important role in determining the structure of the disk corona and emission
features if sufficiently strong.   The irradiation from the
hot corona could also become important in heating the disk if the coronal radiation is very
strong compared to the disk. This may occur at accretion rates when the disk starts to be truncated.
Its affect is relatively unimportant at very low accretion rates when the disk can be
truncated and most of the coronal radiation can escape.  In addition, at very
high accretion rates, the corona irradiation is not as important as the disk radiation itself
although there are indications that the coronal emission can be comparable to the optical emission
in some AGNs. The irradiation can also be important if most of the accretion
energy is released in the corona (e.g. Haardt \& Maraschi 1991; Nakamura \& Osaki 1993; Kawaguchi
et al. 2001), but the nature of the mechanism resulting in such an hypothesis remains unknown.
Therefore, we do not consider the effect of irradiation from the inner corona in our
level of approximation.
With these qualifications, we study the vertical structure in detail, determining the radial
flow owing to the vertically distinct flows in the context of the lower cool disk and upper hot
coronal flows.

\section{Evaporation Model as Applied to AGNs}
\begin{figure}
\plotone{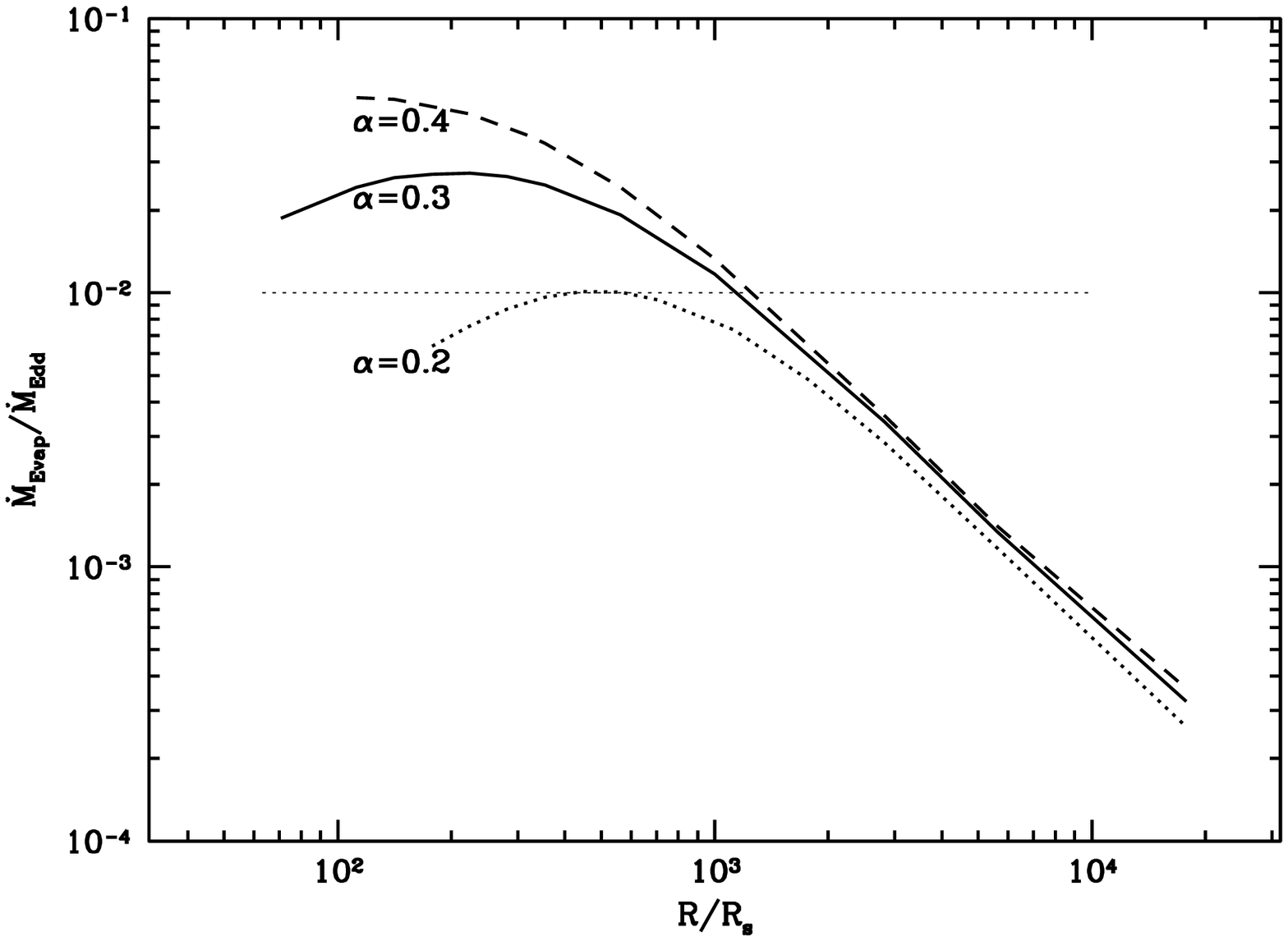}
\caption{\label{f:mdot-r}The mass evaporation rate as a function of distance. For a standard viscosity
parameter, $\alpha=0.3$, the maximal evaporation rate is $0.027\dot M_{\rm Edd}$, taking place at a
distance of $200R_S$ (solid curve). For comparison, the evaporation rate for viscosity parameters,
$\alpha=0.2$ and $\alpha=0.4$, is illustrated by a dotted curve and dashed curve. Intersections of
the thin dashed line with the 3 curves show how the truncation radius varies with the viscosity
parameters, as an example of $\dot m=0.01$.}
\end{figure}

Given the mass of a black hole, $M=10^8M_\odot$, and viscosity parameter, $\alpha=0.3$, we perform
numerical calculations to obtain the vertical structure of the corona for a range of radii.
In these models, the mass evaporation/condensation rate is determined by $\dot m_{z0}=\rho v_{z0}$.
For simplicity, we neglect the possible mass inflow from an outer neighbor since the evaporation
in the outer neighbor region is less efficient than the local evaporation.  As there is no other
mass supply to the corona, the mass flow rate in the corona corresponds to the sum of the
evaporation rate, $\dot M_{\rm evap}= 2\pi R^2 \dot m_{z0}$, from all radii in the disk. This rate
directly determines the relative strength of the corona to the disk as measured by the mass inflow
rate and hence the emergent spectrum.  In Figure \ref{f:mdot-r}, the variation of the mass
evaporation rate with the distance is illustrated. Note that the evaporation rate shown in the
figure is scaled to the Eddington
accretion rate,  $\dot M_{\rm Edd}= 1.39\times 10^{18} M/M_\odot\,{\rm g\,s^{-1}}$, i.e., $\dot
m_{\rm evap}=\dot M_{\rm evap}/\dot M_{\rm Edd}$, and the distance is scaled to the Schwarzschild
radius, $R_S=2GM/c^2$, $r=R/R_S$. It can be seen that the evaporation is negligible at regions far
from the black hole, increasing with decreasing distance until the rate attains a maximum
value, corresponding to $\dot m = 0.027$, at a few hundred Schwarzschild radii.  For smaller radii,
the rate decreases towards the central black hole.
The occurrence of a maximal evaporation rate results from the energy balance in the disk corona. In
the outer region, the evaporation rate increases toward the central black hole since the released
accretion energy increases ($\propto R^{-1}$).  However, with an increase of the number density
towards the inner region, the radiation becomes more efficient (because the bremsstrahlung energy
loss rate is proportional to the square of the number density) and only little gas is heated
sufficiently to evaporate. Thus, the evaporation rate
reaches a maximum value and decreases in the inner region.

Numerical calculations also show that the distribution of evaporation rate with respect to distance,
$\dot m_{\rm evap}(r)$, is independent of the black hole mass, in agreement with our previous finding
(Liu et al. 2002a). Therefore, the model has following predictions as applied to AGNs.

\subsection{Spectral State transition in AGNs}
As described above, the evaporation process diverts the accretion flow from the disk to the
corona.  Consequently, for accretion rates within the disk lower than the maximal evaporation rate,
the disk is completely evaporated within a specific region. This region is filled with coronal gas
(see Figure 2), and the accretion takes place in a geometrically thick flow. On the other hand, for
accretion rates higher than the maximal evaporation rate, the optically thick disk cannot be
significantly depleted at any distance. Thus, the disk extends to the
ISCO with an overlying accreting coronal flow coexisting with the disk, fed by continuous
evaporation.  The variation of the accretion flow geometry with accretion rate is illustrated in
the right panel of Figure 2.

Such an evaporation feature has been successfully applied to interpret
the spectral states observed in BHXRBs, which is believed to be caused by the different accretion
modes. That is, for luminosities greater than the critical luminosity, the accretion is dominated
by the mass flow in the disk with the radiative spectrum described by a multi-color blackbody.
The system is in the so-called soft state. In contrast, for luminosities lower
than the critical value, the accretion is dominated by the corona/ADAF, resulting in a hard state
spectrum. Therefore, the evaporation model predicts a spectral transition from an ADAF to a
disk-dominated accretion flow at a critical luminosity corresponding to the maximal evaporation rate.
 The prediction for AGNs is similar to BHXRBs despite the vast difference in the mass of the
black hole because the evaporation feature is independent of the black hole mass.

The value of the maximal evaporation rate is known to depend on the accretion history. In a low/hard
state, the radiation from the truncated disk is much less than that from the ADAF/corona, and Compton
cooling in the corona due the disk photons can be neglected. Thus, the transition from a low/hard state
to a high/soft state can be appropriately described by the solid curve in the left panel of Figure 3.
That is, the transition takes place at a luminosity of $0.027L_{Edd}$ assuming a standard viscosity,
$\alpha=0.3$. However, the transition from the soft to hard states takes place at a relatively low
accretion rate in comparison to the transition from the hard to soft state.  This is because the
inverse Compton scattering of the disk photons from the ISCO region during the soft state serves as
an additional cooling agent for the coronal gas, which leads to a diminished evaporation rate. In
this case, Compton cooling must be included in the energy equation (\ref{e:energy-t}). The radiation
flux, $F_r=cu$, from the central region as seen by the corona at a distance $R$ is given by
\begin{equation}\label{e:flux}
F_r={L\over 4\pi R^2}{2z\over (R^2+z^2)^{1/2}},
\end{equation}
where $L$ is the luminosity from the central source, which is related to the central accretion rate
by assuming an energy conversion efficiency of 0.1, $L=0.1\dot M c^2$.

Eq.(\ref{e:flux}) reveals that the coronal structure and evaporation rate now depend on the accretion
rate. Based on calculations, it was found that for accretion rates near the values corresponding to
the hard-to-soft transition rate the maximal evaporation rate is less than the accretion rate. Hence,
the disk cannot be truncated at this accretion rate level. However, the evaporation rate increases
with decreasing accretion rates (i.e., decreasing soft photon flux) and once the accretion rate
decreases to $\dot m\approx 0.005$, the maximal evaporation rate equals the accretion rate. This
signifies the onset of disk truncation by evaporation at $\dot m\approx 0.005$ during the decay of
the outburst, as shown by the dashed line in the left panel of Figure 3.  For even lower
accretion rates, the disk is truncated and the accretion flow is dominated by the ADAF/corona.
Therefore, the soft-to-hard state transition takes place at $\dot m \approx 0.005$, which is $\sim 5$
times lower than the accretion rate for the hard-to-soft state transition.  This provides a natural
explanation for the so-called luminosity hysteresis observed in BHXRBs (Meyer-Hofmeister et al. 2005;
Liu et al.  2005).  In AGNs, the different transition luminosities in the soft-to-hard and hard-to-
soft states imply that objects with luminosities in the range $0.005<L/L_{\rm Edd}<0.027$ can be
in either the soft spectral state or hard spectral state. Specifically, if objects are in the
rising phase, their spectra are hard; if they are in the decaying phase, their spectra are soft.

This range in luminosity provides a framework by which objects showing a soft spectrum (big blue
bump and steep X-ray spectrum) can be observed at luminosity levels which are comparable to objects
showing a hard spectrum (without a big blue bump and with a flat X-ray spectrum). Since the
accretion timescale is long compared to the observation timescale, the limited observations do not
distinguish whether an object is in the rising or decaying phase. The model provides a potential
method to determine the phase of variation if the luminosity level lies between the transition
luminosities.  That is, for an object with $0.005<L/L_{\rm Edd}<0.027$ exhibiting a hard spectrum,
it is either a persistent source or, if transient, in the rising phase and a transition to a soft
state is expected to follow, while a soft spectrum indicates the object is in the decaying phase with
a subsequent transition to a hard state expected. The overlap in the Eddington ratio for the soft and hard
spectral states predicted by our model could also lead to a misclassification of AGN sub-types.

To summarize, the evaporation model predicts that objects characterized by luminosities less than
$\sim 0.005L_{\rm Edd}$ are in the low/hard states, whereas those objects with luminosities greater than
$\sim 0.027L_{\rm Edd}$ are in the high/soft state. Within the luminosity range $0.005L_{\rm Edd}$ and
$0.027L_{\rm Edd}$ the objects can be either in the high/soft state or in the low/hard state,
depending on their secular variability. Objects in this luminosity range may undergo variation with
possible transition to another state.

Observationally, spectral state transitions are difficult to detect, although extensive studies
on nearby galaxies (for a review see Ho 2008; Ho 2009) suggest that the galactic nuclei evolve through
distinct states in response to changes in the mass accretion rate. In the sample of Ho (2009),
objects are largely systems in the low or quiescent state, for which the Eddington ratios and
corresponding accretion rates are all below the critical accretion rate predicted by the disk
evaporation model.

Finally, we point out that the quantitative predictions of the evaporation model depend on
the viscosity parameter, $\alpha$, chosen to be 0.3. Larger values of $\alpha$ result in
higher evaporation rates and smaller transition radii. Here, the greater viscous heating leads
to greater conduction of heat to the disk surface, facilitating evaporation.   As a result,
the critical luminosities of the soft-to-hard and the hard-to-soft transition also depend on
$\alpha$.  For example, the critical luminosities of the hard-to-soft transition range from 1\%
to 10\% of the Eddington luminosity for $0.2<\alpha<0.6$ (Qiao \& Liu 2009).

Observational estimates for $\alpha$ in binary star systems have been summarized in King,
Pringle, \& Livio (2007), showing viscosity parameters to be $0.2\la \alpha\la 0.6$ (values are
converted according to the definition $\nu=2/3 \alpha c_s H$ as used here).  However such estimates
for AGN are severely lacking. In a few cases, constraints have been established.  For example, the
investigation of optical variability in AGNs (Starling et al. 2004) suggests a very low viscosity
parameter, $0.01\la\alpha\la 0.03$, if the optical emission is generated in a standard, fully
ionized thin disk and the variability time scale is attributed to the disk thermal timescale.
Nevertheless, these values are noted to be lower limits because shorter time-scales may have been
missed (Starling et al. 2004).

\subsection{The size of broad line region}
Broad lines are a distinguishing feature of type 1 AGNs.  Hints concerning their origin and
nature have been inferred from the empirical relation between the size of broad line region
(BLR) and its bolometric luminosity, $R_{\rm BLR}\propto L_{\rm bol}^{1/2}$, commonly believed
to be a consequence of photoionization of the BLR gas by continuum radiation emitted from the
accreting central regions.  The formation of the BLR has been connected with radiatively
driven winds launched from the disk surface (Murray \& Chiang 1997). If the BLR is, indeed,
associated with the disk through winds, truncation of the disk leads to truncation of the BLR.

An essential feature of the disk evaporation model is the truncation of a disk for accretion
rates below a critical value.  For decreasing rates below this value, the inner radius of the
disk increases.  This property predicts the existence of a region where broad lines would be
excluded in a plane described by the line width and Eddington ratio. The variation of the truncation
radius with respect to Eddington ratio from the numerical calculations is illustrated in Fig.\ref{BLR},
where it is assumed that $L_{\rm bol}/L_{\rm Edd}=\dot m$ for simplicity. If a lower efficiency of
energy conversion is applied to lower accretion rates, the curve is flatter at lower Eddington ratios.
The model predicts that the broad line region lies to the above/right of the truncation curve. The
region lying below the curve corresponds to the region of exclusion.

Three truncation curves are plotted in Fig.\ref{BLR}, showing the dependence of the region of
exclusion on the viscosity and evolutionary stage of an AGN. The solid ($\alpha=0.3$) and dotted
($\alpha=0.2$) lines delineate this region for objects during the rise in outburst.  The difference
between these two curves indicates the sensitivity for the location of the excluded region boundary to
the viscosity parameter. The dash-dotted curve represents the case for objects during decay from
a soft state for a standard viscosity parameter $\alpha=0.3$, which reveals the dependence of the
exclusion region boundary on the evolutionary history.  Therefore, the model predicts a region of
exclusion for the occurrence of broad lines under the dash-dotted curve, if a standard viscous
parameter is adopted, irrespective of the temporal variation of the AGN. In the regime between
the solid and dash-dotted curve, broad lines can be present only for objects decaying from a soft state.

To compare with observations, we take the full sample of 35 reverberation-mapped AGNs (Kaspi et al.
2000; 2005; Peterson et al. 2004) after optical correction of the host-galaxy star light (Bentz et
al. 2006; 2009). The size of BLR, in units of Schwarzschild radius, versus the Eddington ratio is
plotted in Figure \ref{BLR}. The empirical relations of the BLR size with respect to Eddington ratio
are also plotted in the figure for different black hole masses. It can be seen that most of the data
fall into the region between the empirical lines for $M=10^7M_\odot$ and $M=10^9M_\odot$ since this
is the main range of black hole masses of the sample. However, at low luminosities ($L_{\rm bol}/L_{\rm
Edd} \la 0.01$), the BLR size does not continually decrease along the empirical lines.
Therefore, we suggest that the BLR, which is described by the empirical relation, is "truncated" at
low Eddington ratios to a distance corresponding to the inner
edge of the evaporation-truncated disk. With decreasing Eddington ratios, the disk recedes outwards
and hence the inner edge of the BLR increases until the ionization of the supposed BLR is insufficient
for the emission of broad lines.  The data in this small sample is in approximate agreement with this
prediction, though additional data at low luminosities are necessary to confirm it.

 A detailed investigation of the connection between the BLR and disk truncation was also carried
out in an earlier work by Czerny et al (2004) for a very large sample of AGNs.  It was found that the
strong ADAF principle, based on the hypothesis that accretion is via an ADAF whenever an ADAF can
exist, better describes the observational data than the disk evaporation model.  We point out that
a very small value of the viscosity parameter (0.02 for evaporation model and 0.04 for ADAF model)
was required in that study, which is not necessary in the present study. Upon comparison of the two
versions of evaporation model, we find that our detailed numerical calculations of the vertical structure
of disk corona exhibit a dependence of the truncation radius on the viscosity and
magnetic fields than from the approximation-based generalized model (R\`o\.za\`nska \&
Czerny 2000b). 
In particular, a large deviation occurs at Eddington ratio close to the critical transition value.
Specifically, our model reveals that a small change in $\alpha$ results in a large change in the
critical Eddington ratio characterizing disk truncation ($\dot m_{\rm crit} \propto \alpha^{2.34}$),
as illustrated in Figure \ref{BLR} for $\alpha$ varying from 0.2 to 0.3. This sensitivity allows for
the possibility of an inner disk and hence very broad lines at low Eddington ratios by decreasing
$\alpha$ only slightly. In contrast, the weak dependence of truncation radius on $\alpha$ as used
in Czerny et al. (2004) leads to a very small value for $\alpha$ in order that broad lines be present
at low Eddington ratios.  With our detailed numerical model, we expect that the observational sample
adopted by Czerny et al. (2004) can also be explained provided that $\alpha\sim 0.1$.  This implies
that it is unnecessary to assume very small $\alpha$ when the detailed disk evaporation model is
used to explain the broad line emissions.  In this sense, the disk evaporation model is not as
restrictive as compared to models based on the strong ADAF principle.

We note
that the theoretical curves are nearly parallel at low Eddington ratios and that the truncation radii are not as sensitive to $\alpha$ in this regime as compared to Eddington ratios of $\sim 0.01$.
For a normal viscosity value and assuming an appropriate ionization, Fig.\ref{BLR} shows that
 the broad line region can be present at  Eddington ratios as low
as $\sim 0.001$ .
However, the size of BLR is not as small as described by the empirical law since the evaporation
truncates the inner part or all of the BLR at low Eddington ratios.
Therefore, the disk evaporation model predicts two regimes in which the broad lines are
"narrower": one corresponding to very low $L/L_{\rm Edd}$ where the disk is truncated at large
distances associated with a large broad line region (i.e. LLAGNs, for a review see Ho 2008);
the other is at very high $L/L_{\rm Edd}$ where a full disk is present, but its partially ionized
region is at large distances (such as the NLS1s). Caution is necessary because there exist
uncertainties in the observational data and the adopted parameters in spectral fitting, both of
which could lead to large error bars in the derived accretion rate and truncation radius.

\subsection{The lower limit to the luminosity for the existence of a broad line region and the occurrence of ``True Seyfert 2'' Galaxies}
For very low accretion rates the disk is truncated at large distances by evaporation. For
instance, the disk is truncated at 6600$R_S$ at $\dot m=0.001$ (in the case of standard viscosity
$\alpha=0.3$ and no magnetic fields).  If regions at such large distances can still emit broad
lines, the emission line would be quite narrow.  The FWHM of emission lines, as estimated by the
Keplerian velocity with a correction factor $2/\sqrt{3}$ (which accounts for velocities in three
dimensions and the full width corresponding to twice the velocity dispersion), is
given by
\begin{equation}
FWHM={2\over \sqrt{3}}\sqrt{GM\over R_{\rm BLR}}=\sqrt{2c^2\over 3R_{\rm BLR}/R_S}
\end{equation}
Thus, at $R_{\rm BLR}=6600R_S$, the FWHM of emission line is $\sim 3000$ km s$^{-1}$.
If we take 2000 km s$^{-1}$ as the critical linewidth
between broad line AGNs and narrow line Seyfert 1 galaxies, we obtain an inner radius of the BLR as
$15,000R_s$.  To truncate the disk to this distance, the accretion rate should be $4\times
10^{-4}$ in units of the Eddington rate as derived from the disk evaporation model.  At even lower
accretion rates, the disk is truncated at a larger radius and hence the BLR would no longer be
recognizable.   Therefore, the disk corona evaporation
model predicts that the lower limit to the luminosity for the inference of a BLR is $\sim
0.0004-0.001 L_{Edd}$.

The lower limit to the luminosity of AGNs characterized by a BLR implies that the orientation
based unification model is not applicable to LLAGNs. Instead, objects with Eddington ratios
lower than $\sim 0.001$ may intrinsically lack a BLR rather than simply be obscured by a torus.  In
other words, the model predicts that there exist so-called ``true Seyfert 2'' galaxies. This is
supported by the observations of Bianchi et al. (2008), who discovered an unobscured Seyfert 2
galaxy, NGC 3147, without a BLR. The Eddington ratio for this object ranges roughly between $8\times
10^{-5}$ and $2\times 10^{-4}$. Similar discoveries of naked AGNs, characterized by the absence of
a BLR accompanied by strong continuum emission and strong variability in the optical band and/or
without significant intrinsic absorption in X-rays, indicate that these sources genuinely lack a
BLR (see Hawkins 2004; Gliozzi et al. 2007).

Observations of Seyfert 2 galaxies (Sy2s) also support the existence of ``true Seyfert 2'' galaxies.
Of the brightest  Sy2s only $\sim 50\%$ exhibit the presence of hidden BLRs in their polarized
optical spectra (Tran 2001, 2003; Moran 2006).  The luminosities of Seyfert 2 galaxies
without a hidden BLR have systematically lower luminosity/Eddington ratios (Lumsden \& Alexander
2001; Gu \& Huang 2002; Tran 2001, 2003; Moran 2006; Bian \& Gu 2007) than Sy2s with hidden BLR
(obscured by inclination effects). This implies that ``true Seyfert 2s'' exist at
low luminosities. Quantitative investigations of the non-BLR yield an estimate for the
threshold of the Eddington ratio of $10^{-1.37}$ for Sy2s (Bian \& Gu 2007) and $10^{-2}$ for naked
AGNs (Gliozzi et al. 2007), both of which are higher than the theoretical prediction.  We note,
however, that large uncertainties exist in estimating the bolometric luminosity from the
optical or X-ray luminosity (e.g., the bolometric luminosity in Bian \& Gu (2006) should be 1.5dex
lower if based on the method used in Gliozzi et al. (2007)), and uncertainties are present in the
determination of the black hole mass, both contributing to apparent deviations from the theoretical
prediction.

\section{Discussion and Remaining Issues}

We have shown that a model developed for BHXRBs based on a disk-corona interaction provides
a promising interpretative framework for explaining the characteristic observational features of
AGNs. The evaporation of gas leads to truncation of the disk for accretion rates less than a maximal
evaporation rate.  This characteristic naturally accounts for the distinct states of AGN, viz., a
soft spectrum with big blue bump at high luminosities and a hard spectrum at low luminosities.
Due to the existence of differing critical mass flow rates describing spectral transitions
from the hard state to the soft state and the soft state to the hard state, AGNs characterized
by the same Eddington ratio within the critical luminosity interval can be in either state.

Associating the characteristic scale of the broad line region, $R_{BLR}$, with the disk
truncation radius for LLAGNs, the model predicts that the region excluding the
presence of broad lines increases with decreasing luminosity for luminosities less than
about $0.005 -  0.027 L_{Edd}$ for an $\alpha$ viscosity parameter of 0.3. Hence, $R_{BLR}$
departs from the relation $R_{BLR} \propto L^{1/2}$ at these luminosity levels.
As a consequence, broad emission
lines are only expected for AGNs characterized by mass accretion rates above a given value, if
the broad line region is associated with an optically thick accretion disk. Thus, true
Seyfert 2 galaxies are expected at sufficiently low luminosities.

As an additional consequence of the disk truncation, a thermal red bump is expected to be
present at low Eddington ratios. This expectation is in qualitative agreement with the presence
of the so called ``big red bump'' in LLAGNs (see Ho 1999; 2008). A detailed study on the disk
truncation and a comparison with the observational properties of the big red bump is reserved
for a future paper.

Although the model's basic features provide theoretical support for the hypothesis that the accretion
process for supermassive black holes in AGNs is similar to the process for stellar mass black holes
in BHXRBs, several observational differences exist between these two types of systems.  In the
following, we discuss these differences and their possible implications for the model.

\subsection{Strong coronal flow in AGNs}

As the evaporation is saturated to the maximal rate at high accretion rates (Eddington ratios), the
accretion flow in the corona cannot increase with the accretion rate after a transition to the
soft state. Thus, the accretion flow within the disk (rather than corona) increases with increasing
accretion rate. With efficient Compton scattering, the corona component could be even weaker at
high accretion rates because the increased disk radiation leads to strong Compton cooling, resulting
in the condensation of coronal gas and the reduction of the optical depth in the corona. If we
adopt the maximal evaporation rate without Compton cooling as an upper limit for the corona flow,
we obtain an upper limit on the ratio of the mass inflow rates through the corona and disk as
\begin{equation}
{\dot m_c\over \dot m_d}={1\over {\dot m\over 0.027}-1},
\end{equation}
which is also an upper limit to the ratio of the coronal luminosity to the disk luminosity of
a source in the high/soft state.  For an accretion rate of 0.1 the coronal luminosity is no
more than 37\% of the disk luminosity and is less than $\sim 3\%$ of the disk luminosity if
the system accretes at the Eddington rate. This implies that the spectral index between optical
and X-rays, $\alpha_{\rm ox}$, is very large in luminous AGNs.

However, observations reveal that the X-ray luminosity is comparable with the optical luminosity in
some HLAGNs, indicating a strong corona in AGNs (which is similar to the very high state in BHXRBs).
 To produce such strong X-ray
radiation in luminous AGNs, the frictional heating (in the model) is insufficient. To address
this conundrum, it has been suggested that a large fraction of accretion energy is transported
from the disk to the corona, perhaps due to a magnetic buoyant instability (Liu et al. 2002b;
2003; Cao 2009).  Alternatively, the viscosity parameter $\alpha$ in these HLAGNs may be large,
supporting a strong corona by means of efficient evaporation.  Such a possibility could be a
consequence of a larger magnetic Prandtl number in AGNs than in BHXRBs (see Lesur \& Longaretti 2007).

\subsection{Additional phenomenology}

In addition to their apparent strong coronae, AGNs exhibit other observational features which differ
from BHXRBs. For example, among the luminous AGNs $\sim 10\%$ are radio-loud QSOs, while jets in
BHXRBs appear to quench whenever the objects transit to the high/soft state. This difference may
reflect that jets in the high state of BHXRBs are weak or the properties of the magnetic field in
the corona, related to the magnetic Prandtl number, are modified (see above) to facilitate jet
formation in AGNs.

We also note that soft X-ray excesses are seen in AGNs at high Eddington ratios, but not in BHXRBs.
Such excesses appear to be correlated with the presence of the big blue bump.  The origin of this
excess is unknown, but it has been interpreted in terms of Comptonization in the disk (Kawaguchi et
al. 2001) or resulting from a depression of the continuum at X-ray energies in the range 0.7-5 keV
by relativistically broadened line transitions of O VII and O VIII (Gierlinski \& Done 2004). Its
restriction to AGNs remains to be clarified.

The environment of a galactic central region can significantly differ from that for a stellar
mass black hole in an X-ray binary system. For example, significant high temperature diffuse gas has
been observed in the Galactic Center, with Sgr A$^*$ accreting at very low Eddington ratios,
which may also be present in some LLAGNs. Such hot gas would affect the boundary condition on
the disk. However, in the case of hot gas joining the evaporating gas in the corona, the
corona is not changed significantly. This is a consequence of the fact that the evaporation of
the disk is depressed by the directly inflowing gas due to the pressure and energy balance between
the disk and the corona. If the density in the corona is too high, the efficient cooling to the
corona can even lead to the coronal gas condensing to the disk. Consequently, the corona flow
remains similar to the case of evaporation-fed corona.  In other words, for a given accretion rate,
independent of whether the disk is fed at the outer boundary, or the disk (corona) is
partially fed in the form of cold (hot) gas, the accretion flows through the disk and through
the corona in regions of a few hundred Schwarzschild radii are expected to be similar.

Finally, the disk instability model may not give rise to large amplitude outbursts (and hence
transitions between the low/hard state and the high/soft state) due to the ionization
instability in AGNs as in BHXRBs because the $\alpha$ parameter in the cold state may not
significantly differ from its value in the hot state (see Menou \& Quataert 2001). In this
case the variability would be of small amplitude and state transitions would be less dramatic.
AGNs would therefore primarily reflect the mass inflow rate from its surrounding region rather
than a temporal variation in a non-steady disk. On the other hand, gravitational instabilities
may operate in AGNs to give rise to more significant time dependent accretion.

In the future, we plan to carry out additional investigations to quantitatively confront the disk
corona model with the detailed observations of AGNs, with comparisons based on a two-temperature
corona model (Liu et al. 2002a) incorporating the effect of magnetic fields (Qian et al. 2007) and
viscosity (Qiao \& Liu 2009).  In addition, a study of their differences with respect to BHXRBs
will be explored to examine whether they are one of degree rather than of kind.

\acknowledgments
We are grateful to Friedrich Meyer and Emmi Meyer-Hofmeister for stimulating discussion and
comments on the manuscript. We thank Luis Ho for his careful reading of a preliminary version
of the manuscript and very helpful suggestions. Bozena Czerny is thanked for the early discussion. Finally, the referee is thanked for the detailed
comments which improved the clarity of the paper. Financial support for this work is provided by the
National Natural Science Foundation of China (grants 10533050 and 10773028) and by the National
Basic Research Program of China-973 Program 2009CB824800. In addition, R.E.T. acknowledges support
from the Theoretical Institute for Advanced Research in Astrophysics (TIARA) in the Academia Sinica
Institute of Astronomy \& Astrophysics.

\begin{figure}
\plottwo{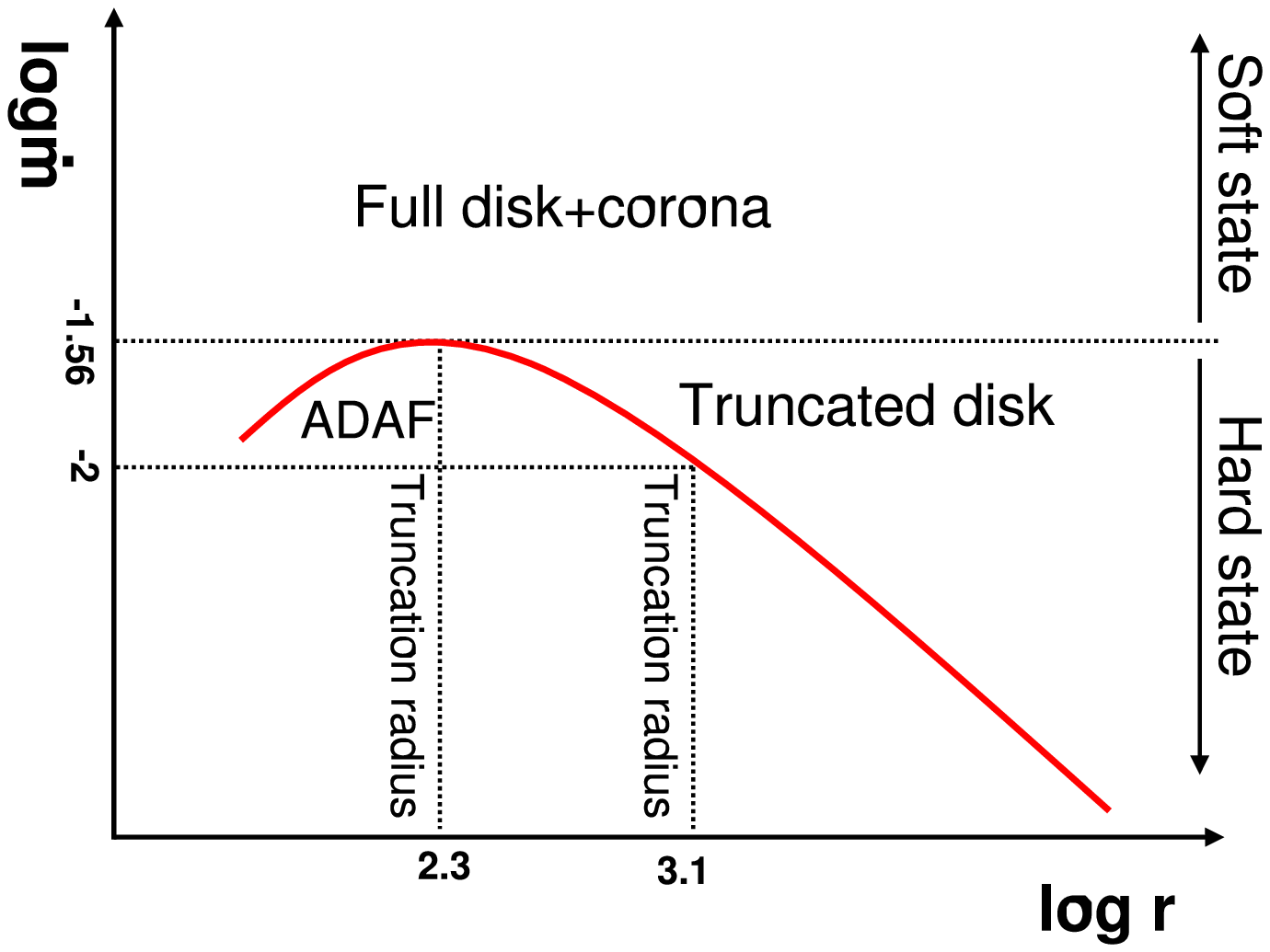}{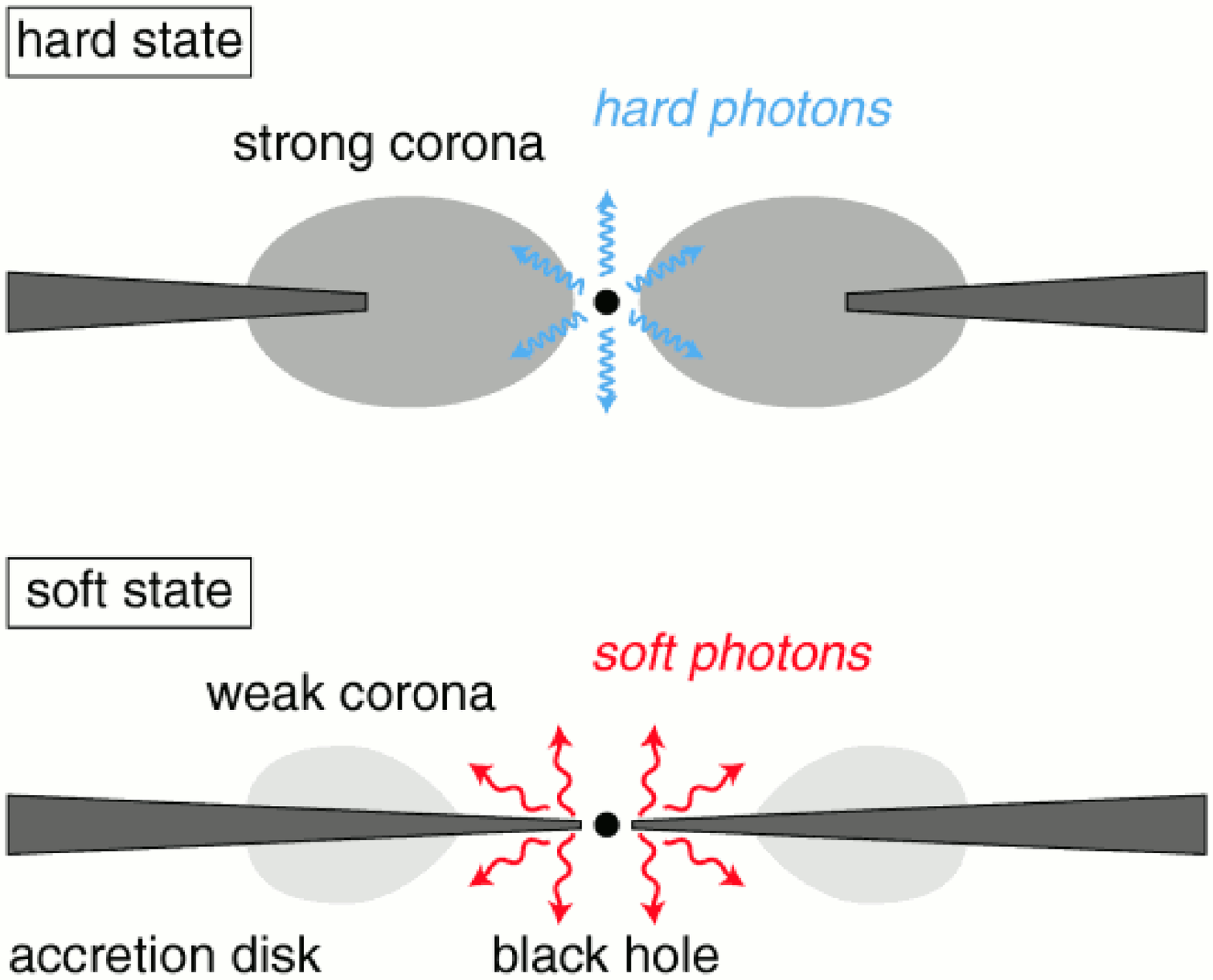}
\caption{\label{truncation} A schematic description of disk truncation and the spectral transition
illustrating its dependence on the mass accretion rate. The evaporation rate is illustrated as a function
of distance in the left panel.  At low accretion rates, the disk is truncated by evaporation due
to insufficient mass supply at distances determined by the evaporation curve. The accretion flow in
the inner region takes place via a corona/ADAF. When the accretion rate exceeds the maximum, which
is 0.027 times the Eddington rate in the case for which Compton cooling is unimportant and a
standard viscosity parameter ($\alpha=0.3$), evaporation cannot evacuate any disk region since the mass
flow in the disk is greater than the evaporation rate. Thus, the disk extends to the ISCO and dominates
the accretion flow.  The corona overlying the disk becomes very weak. The figure implies that objects
at high accretion rate/luminosity are in a soft spectral state, while below this accretion rate/luminosity,
objects are in a hard spectral state with a state transition occurs at an accretion rate corresponding
to the maximal evaporation rate. The corresponding accretion geometry is plotted in the right panel.}
\end{figure}

\begin{figure}
\plottwo{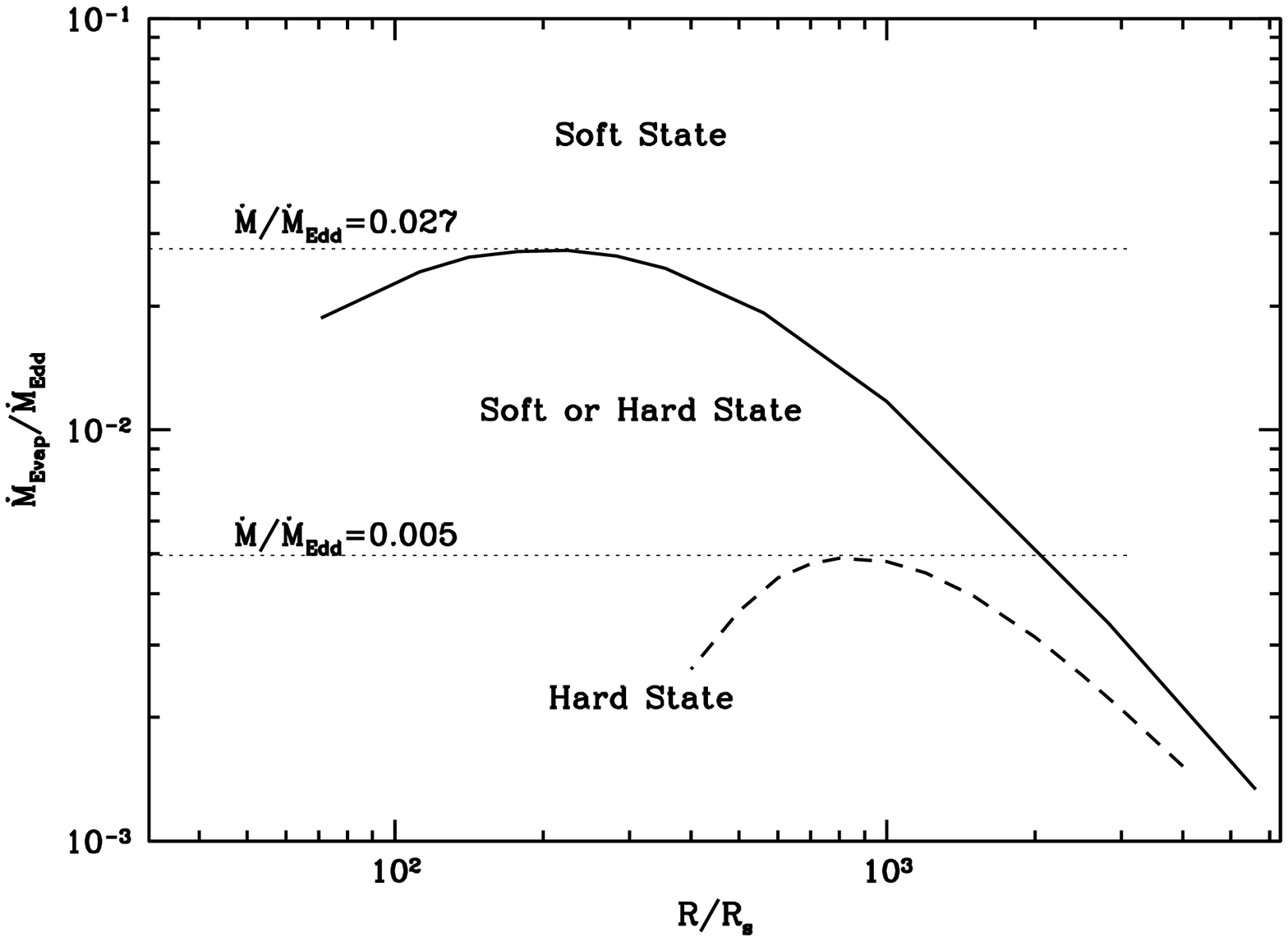}{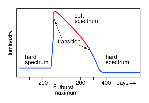}
\caption{\label{trans}Left panel: Model prediction of the spectral states of AGNs and dependence
on accretion history. If any inner disk is absent, the maximal evaporation rate is $\dot m_{\rm evap,
max}=0.027$. The state transition takes place at 0.027 times the Eddington rate, as shown in Figure 2.
If an inner disk dominates the accretion flow, the evaporation is greatly reduced by Compton cooling of
the corona by soft photons from the disk. A transition occurs at an
accretion rate of 0.005 times the Eddington value.  This implies that the spectral transition from a
soft to hard state takes place at $\dot m= 0.005$ and from hard to soft at  $\dot m= 0.027$.
Specifically, for an object undergoing an increase in luminosity from a quiescent state, the
spectrum is hard until the accretion rate reaches 0.027 times the Eddington value; whilst during the
decay phase from a soft state, the transition to a hard state takes place only when the accretion rate
decreases to a lower value, $\dot m=0.005$ where the disk is again truncated by evaporation. This
description provides an interpretation for the hysteresis observed in BHXRBs. For AGNs, the consequence is,
for accretion rates $0.005\la\dot m\la 0.027$, the accretion flow can be dominated either by a disk
or by an ADAF. Therefore, the model predicts that objects with an Eddington ratio in the range between
0.005 and 0.027 can be in either the soft state or hard state, depending on its accretion/evolutionary
history. Right panel: Schematic light curve of a representative outburst for a typical BHXRB. The
transition luminosity from the hard to soft state is higher than that from the soft to hard state,
in agreement with the model. }
\end{figure}

\begin{figure}
\plotone{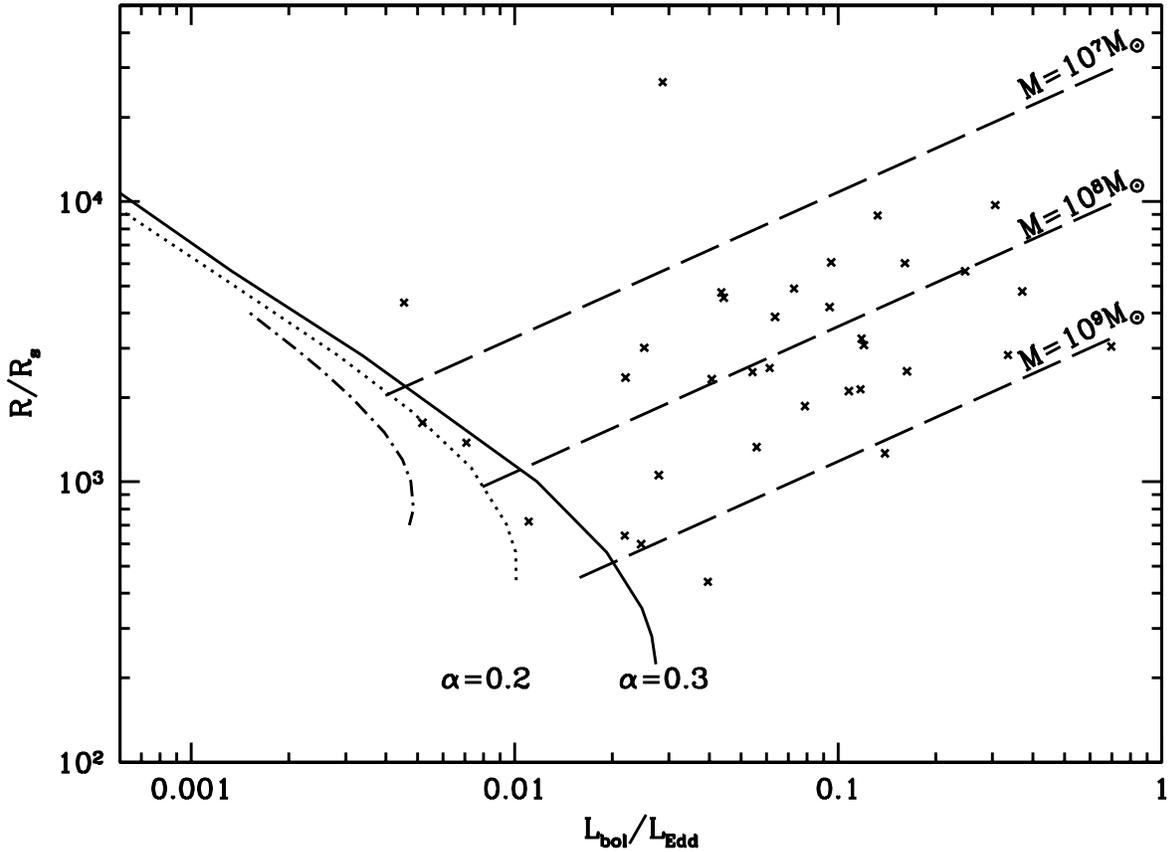}
\caption{\label{BLR}The size of BLR vs the Eddington ratio. The data are taken from
reverberation mapping of Peterson et al. (2004), Kaspi et al. (2005), and Bentz et al. (2006, 2009).
The empirical relations of the size of BLR and Eddington ratio for a black hole mass of
$M=10^7M_\odot, 10^8M_\odot$, and $10^9M_\odot$ are plotted as long dashed lines.  The curves
indicate the truncation radius of the disk as a function of Eddington ratio for $\alpha=0.3$ (solid
line) and $\alpha=0.2$ (dotted line) without Compton cooling and with Compton cooling (dash-dotted
line) for the standard viscosity parameter $\alpha=0.3$. The model predicts an absence of the BLR below
the truncation curve because the inner disk is absent in this regime. The broad line can be present
in between the solid curve and dash-dotted curve if the object evolves from a soft state. In contrast,
it is absent if it evolves from a hard state.  The data show that the size of BLR increases with
the Eddington ratio along the empirical lines at high luminosities/Eddington ratio, while at very
low Eddington ratios, the BLR appears to be ``cut out'' along the model prediction curve. }
\end{figure}
\end{document}